\documentclass[preprint,showpacs,preprintnumbers,amsmath,amssymb,12 pt.,one-column,double-spaced]{revtex4-1}

\usepackage{graphicx}
\usepackage{dcolumn}% Align table columns on decimal point
\usepackage{bm}
\usepackage{color}

\begin{document}

\title{Persistent circular currents of exciton-polaritons in cylindrical pillar microcavities}

\author{V.A.~Lukoshkin$^{1,2}$}

\author{V.K.~Kalevich$^{1,2}$}

\author{M.M.~Afanasiev$^{1,2}$}

\author{K.V.~Kavokin$^{1,2}$}

\author{Z.~Hatzopoulos$^{3}$}

\author{P.G.~Savvidis$^{3,4,5}$}

\author{E.S.~Sedov$^{6,7}$}

\author{A.V.~Kavokin$^{1,6,8}$}

\affiliation{$^1$Spin Optics Laboratory, Saint-Petersburg State
University, 1~Ulianovskaya, St-Petersburg 198504, Russia}

\affiliation{$^2$Ioffe Institute, Russian Academy of Sciences, 26~Politechnicheskaya, St-Petersburg 194021, Russia}

\affiliation{$^3$IESL-FORTH, P.O. Box 1527, Heraklion 71110, Greece}

\affiliation{$^4$Department of Materials Science and Technology, University of Crete, Heraklion 71003, Greece}

\affiliation{$^5$National Research University for Information Technology, Mechanics and Optics (ITMO), St-Petersburg 197101, Russia}

\affiliation{$^6$School of Physics and Astronomy, University of Southampton, Highfield, Southampton SO171BJ, UK}

\affiliation{$^7$Department of Physics and Applied Mathematics,
Vladimir State University named after A.G. and N.G. Stoletovs,
87~Gorky str., Vladimir 600000, Russia}

\affiliation{$^8$CNR-SPIN, Viale del Politecnico 1, I-00133, Rome, Italy}

\begin{abstract}

We have experimentally observed an eddy current of exciton
polaritons arising in a cylindrical GaAs/AlGaAs pillar microcavity under the nonresonant optical pumping. The polariton current
manifests itself in a Mach-Zehnder interferometry image as a characteristic spiral that occurs due to the interference of the light emitted by an
exciton-polariton condensate with a spherical wave artificially
shaped from the emission of the same condensate. We have
experimentally observed the condensates with the topological charges $m=+1$, $m=-1$
and $m=-2$. The interference pattern corresponding to the $m=-2$ current represents the twin spiral emerging from the center of the micropillar.

\end{abstract}

\pacs{71.36.+c, 73.20.Mf, 78.45.+h, 78.67.-n}

%\keywords{polariton, condensate, semiconductor microcavity, exciton}

\maketitle

Exciton-polaritons are superposition quasipartcles formed in the
strong exciton-photon coupling regime in various semiconductor
structures. Since 1992, a particular attention has been attracted
to exciton-polaritons in semiconductor microcavities~\cite{Weisb}.
Several fascinating effects linked with the bosonic nature of
exciton-polaritons have been observed, including e.g. the
stumulated scattering~\cite{Savvidis2000}, Bose-Einstein
condensation and polariton
lasing~\cite{KasperNature2006,Balili2007}. The formation of
bosonic condensates of exciton-polaritons that is at the heart of
polariton lasing manifests itself by the spontaneous emission of a
coherent and monochromatic light by a
microcavity~\cite{Imam1996,Christopoul2007}. Being formed by
weakly interacting bosonic quasiparticles polariton condensates
exhibit some characteristic features of quantum fluids including the quantized vortices, similar to those observed in
superconductors and in the superfluid helium~\cite{Laggett}.

The spontaneous formation of vortex-antivortex pairs in
non-resonantly pumped polariton condensates has been observed in
Refs.~\onlinecite{Lag2008,Nard,Roum}. The spontaneous generation
of persistent circular currents is expected in polariton liquids
confined in annular traps~\cite{Nalitov2017}. Such currents
represent a significant interest from the fundamental point of
view and may be promissing for applications in quantum
interferometers and gyroscopes. Circular persistent currents are
being studied in cold atomic condensates in optical traps
generated by Laguerre-Gaussian light modes that transfer their
angular momentum to the atomic condensate~\cite{Rama,Moulder}.

In this context, exciton-polariton condensates offer an advantage
of relatively easy tailoring of their shapes due to the repulsion
of the coherent condensate fraction from the excitonic reservoir
generated by a spatially inhomogeneous optical
pumping~\cite{Tosi2012}. Moreover, chemical etching of
planar semiconductor microcavities allows for formation of pillars
providing a deep confining potential for
excitons-polaritons~\cite{PRL1000474012008,KVKJAP,KVK}. The theory of
polariton quantum liquids is now well-developed~\cite{Berloff}.
The experimental methods of generation and detection of polariton
circular currents are discussed in Refs.\,\cite{Egor,Ostr}.  
Nevertheless, to our knowledge, the direct experimental evidence
for formation of persistent circular currents in ring-shape
polariton condensates has not been reported so far.

In our previous works we have demonstrated experimentally the formation of the ring-shaped polariton condensates in cylindrical pillars of different diameters~\cite{KVK,KVKJAP}. In this Letter, we report on the experimental observation of persistent circular polariton currents with different projections of the orbital momentum to the axis of the structure that correspond to different topological charges $ m $ in cylindrical semiconductor micropillars under non-resonant optical excitation with Gaussian beams at the center of the pillar.

A Bose-Einstein condensate represents a macroscopically occupied
single quantum state~\cite{Yang}. Herewith the multiparticle
quantum system can be described by a single complex wavefunction
$\Psi = \Psi(\mathbf{r},t)\exp[i\Phi (\mathbf{r},t)]$ possessing a well-defined
quantum mechanical phase $ \Phi$. Photons, emitted by the condensate,
keep information about the condensate phase. This opens the possibility to
investigate coherent properties of the exciton-polariton
condensate using optical interferometry methods. For this purpose
we use the Mach-Zehnder interferometer with the reference
spherical wave.

\begin{figure}[t]
    \includegraphics[width=1.0\linewidth]{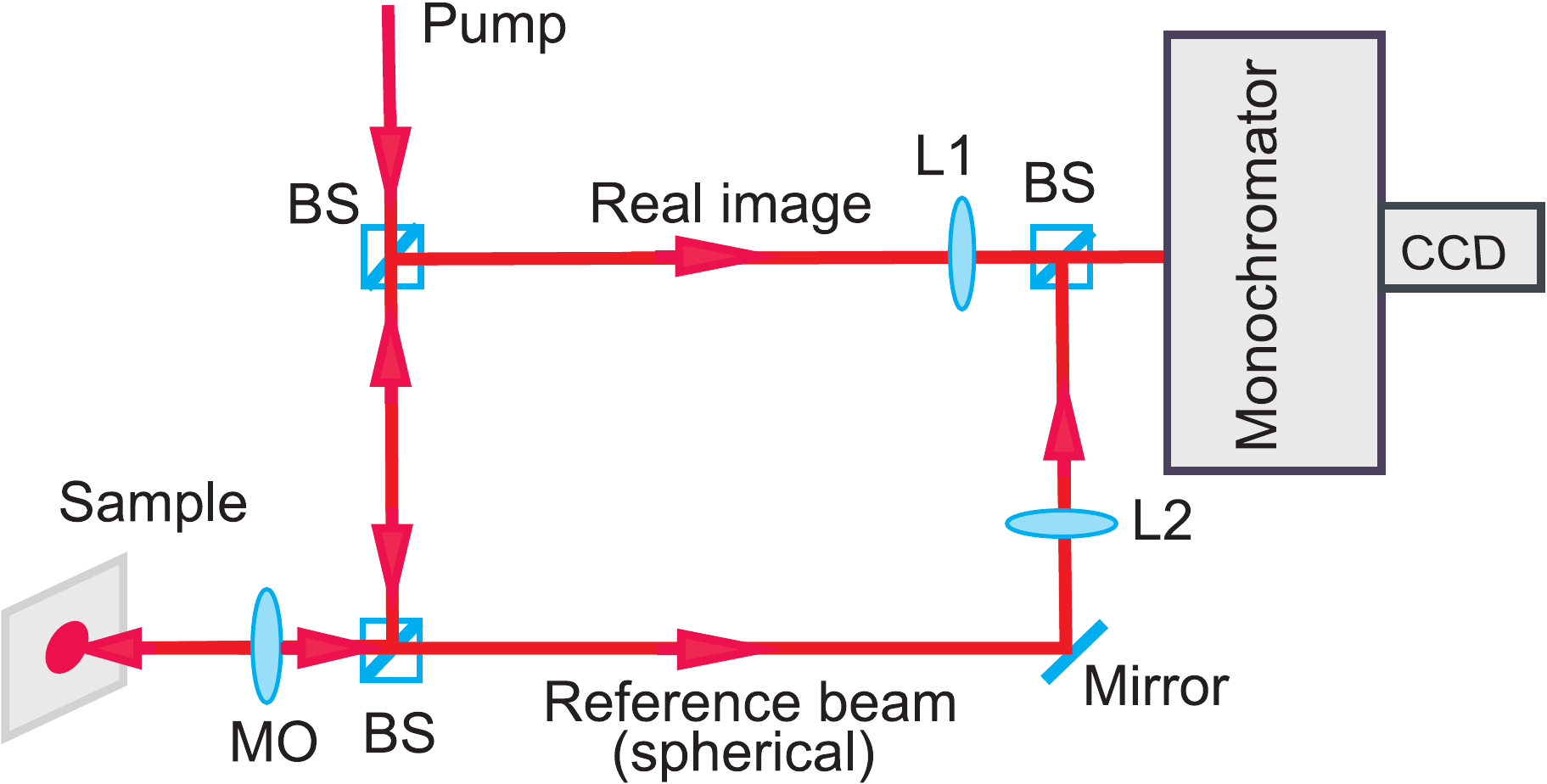}
    \caption{\label{Fig1}
        (Color online)
        Schematic of the interferometer setup for the investigation of coherent properties of the exciton-polariton condensate in a micropillar.
        The microscope objective (MO) is used for focusing the excitation laser beam to a 2$\,\mu$m spot on the sample surface and for collecting photoluminescence.
        The real-space image of the pillar is projected on the entrance slit of the 50-cm monochromator by the lens L1 (focal length $F=700$\,mm).
        The reference beam was converted to the spherical wave using the lens L2 with $F=100$\,mm.
        }
    \label{Fig1}
\end{figure}

The general scheme of our setup is presented in
Fig.~\ref{Fig1}. The detector block is based on the Mach-Zehnder interferometer. The exiton-polariton
condensate in the pillar microcavity was nonresonantly pumped by
a cw Ti:sapphire laser tuned to the local minimum of the upper
stop-band of the distributed Bragg reflector (about $110$\,meV
above the low polariton branch minimum). Because of nonresonant excitation we do not imprint any phase distribution in the condensate directly with the pump laser. The linearly polarised
laser beam was focused to a 2\,$\mu$m spot by a microscope
objective (MO) of the numerical aperture of 0.42. The same
MO was used to collect the photoluminescence (PL) from the
condensate. The parallel PL beam after MO passes through the beamsplitter (BS) and splits into two beams. The upper (in the scheme in Fig.~\ref{Fig1}) beam, passing
through the lense L1 produces an enlarged ($\approx 6$\,mm) real
image of the condensate at the entrance slit of the 50-cm
monochromator. The lense L2 transforms the lower beam to the
spherical wave that is used as a reference. Both beams
overlap at the output BS of the interferometer, and their
superposition is projected on the entrance slit of the
monochromator. The interferogram (or the near-field image of the condensate if the reference beam is blocked) was recorded by the cooled
CCD-camera at the output of the monochromator. 

All experiments were performed at normal incidence of the
excitation beam on the sample. A cut-off interference filter was
installed in front of the monochromator entrance slit to suppress
the excitation laser radiation scattered from the pillar surface.
The sample under study was kept in the helium-flow cryostat at
$T=3.5$\,K.

We have examined a set of cylindrical pillars with a diameter of
25\,$\mu$m that were etched from a planar $5\lambda /2$ AlGaAs
distributed Bragg reflector microcavity with the measured quality
factor of $Q=16000$. Four sets of three 10\,nm GaAs quantum wells
are placed at the antinodes of the cavity electric field to
maximize the exciton-photon coupling~\cite{Tsotsis2012}. The
microcavity wedge allowed scanning across the sample to set the
detuning energy $\delta=E_{C}-E_{X}$, where $E_{C}$ and $E_{X}$
are energies of the cavity mode and of the heavy-hole exciton at
zero in-plane wave vector. The studied pillars are
characterized by a negative photon-exciton detuning $\delta = -
(0.5 \div 3.5)$\,meV. Within this range, we have found no strong
qualitative variation of the observed effects.

\begin{figure}[t]
    \includegraphics[width= 0.9\linewidth]{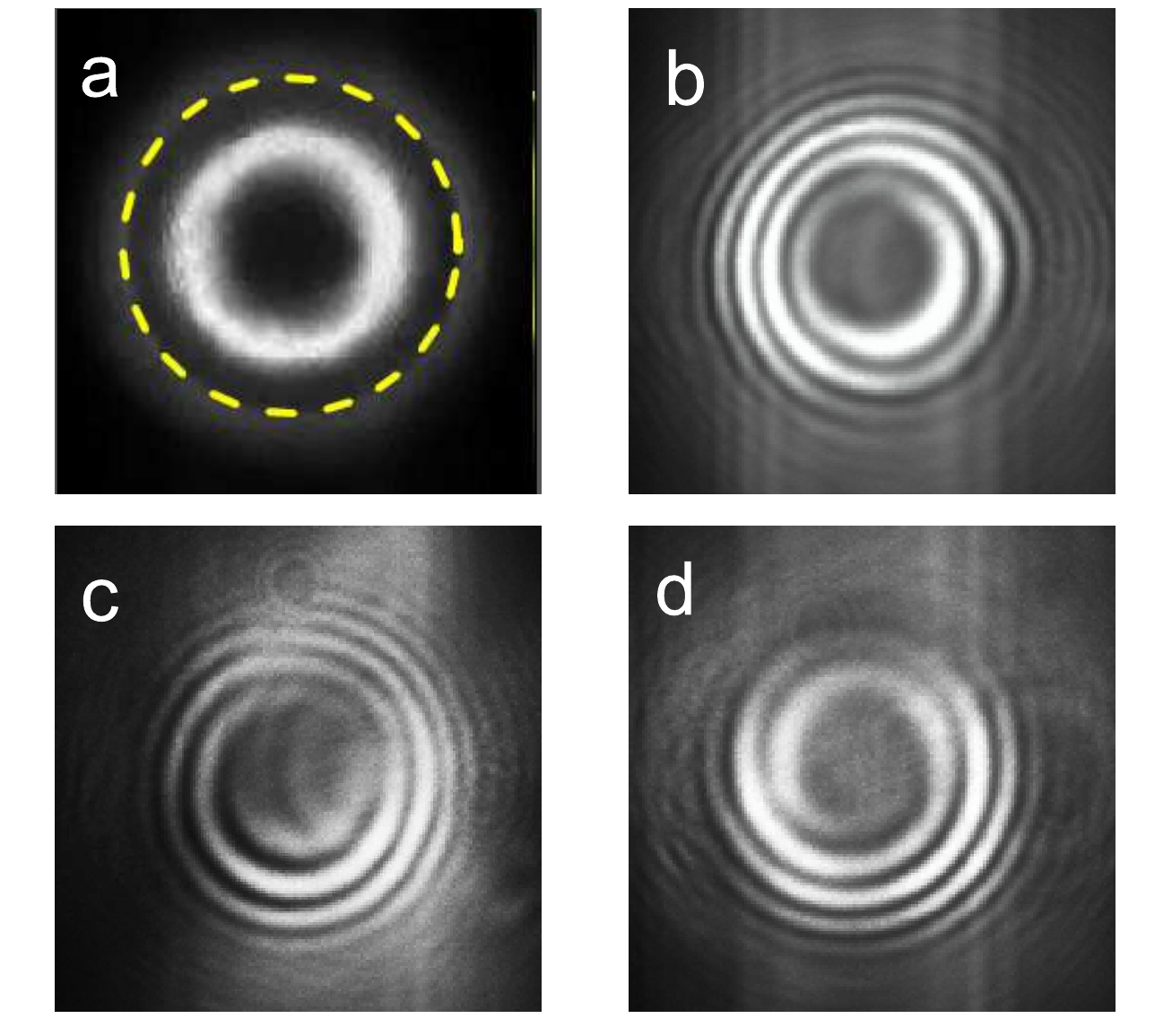}
    \caption{
        \label{Fig2}
        The real-space image of the exciton-polariton condensate measured in the cylindrical micropillar
        of a diameter of 25\,$\mu$m under the nonresonant pump in the center of the pillar (a).
        The experimental interferograms of the exciton-polariton condensates with the topological charges $ m = 0$  (b), $ m = +1$ (c) and $ m = -1$ (d).
        The energy of the excitation quanta is $\ h\nu_{\textrm{exc}}=1.664$\,eV,
        the pump power is $P \approx 1.5P_\textrm{th}$,
        the temperature is $T=3.5$\,K.
        The dashed circumference in the panel (a) indicates the edge of the pillar. All plots are normalized to the maximum intensity.}
\end{figure}

The real-space image of the condensate and its interferogram measured when the pump beam was focused in the pillar center are
shown in Figs.\,\ref{Fig2}a and \ref{Fig2}b,
respectively, for the pump power $P \approx 1.5P_\textrm{th}$,
where $P_\textrm{th} = 2.8$\,mW is the condensate threshold. As
seen in Fig.\,\ref{Fig2}a, the condensate has the shape
of a symmetric ring with the diameter ($\approx 16$\,$\mu$m)
strongly exceeding the size of the excitation spot ($\approx
2$\,$\mu$m), in full agreement with our earlier
observations~\cite{KVK,KVKJAP} in what concerns the dependence of the condensate shape on the diameter of the pillar, the intensity and the position of the pumping beam. The interferogram in Fig.\,\ref{Fig2}b has the form of a set of concentric
rings, evidencing the absence of the non-trivial topological charge in
the condensate, $m = 0$. It is worth to note that the nearly perfect
shape of the interference rings, as well as their concentricity,
certify the sphericity of the wave formed in the reference arm of
our interferometer.

A small displacement (less than 1\,$\mu$m, see details below) of the pumping spot from the position where the image with $m = 0$ in
Fig.\,\ref{Fig2}b was recorded, leads to a dramatic
change of the interferogram. In particular, the interference
pattern can be transformed into the single spiral turning
counter-clockwise (Fig.\,\ref{Fig2}c) or clockwise
(Fig.\,\ref{Fig2}d). The spiral shape of the
interference fringes indicates that the dependence of the condensate phase $\Phi$ on the azimuth angle $\theta$ is close to linear: ${\Phi = m
\theta}$. The observation of one-thread spirals with opposite
helicities evidence occurrence of the non-trivial topological charge of
the condensate, equal to unity with both possible signs: $m = +1$
in Fig.\,\ref{Fig2}c and $m = -1$ in
Fig.\,\ref{Fig2}d. For our ring-shaped condensates,
the appearance of a non-zero angular momentum indicates that a circular
current of polaritons flows around the ring. We would like to
stress that once these current states emerge they are quite stable.
The interference images remain unchanged for minutes that is many orders of magnitude longer that the polariton lifetime (about
10~ps). This brings us to the conclusion that we observe persistent circular currents of Bose-condensed exciton-polaritons trapped in the cylindrical pillar.

All interferograms in Fig.~\ref{Fig2} were obtained at
the same experimental conditions except for the position of the
excitation spot, which was slightly different in different experiments. The shift of the
excitation spot was certainly less than 2\,$\mu$m, which is the
positioning accuracy of our micrometric mechanical X-Y translation
stage that performs movement of the sample with respect to the
excitation beam. In our opinion, the real displacement of the
pumping spot away from the pillar center was even much less
than 1\,$\mu$m. This estimate is supported by the fact that the
near field condensate image, which is extremely sensitive to this
displacement~\cite{KVK,KVKJAP}, being measured for each
specific interferogram, was unchanged, coinciding with the image
in Fig.\,\ref{Fig2}a. This observation suggests a very strong
sensitivity of the condensate current states to the effective
potential landscape created by the reservoir of incoherent
excitons, by the pillar boundary and by various inhomogenities and
defects in microcavity.

Due to a very short polariton lifetime, the polariton condensate
exists as long as the optical excitation is present and disappears soon after switching off the pump. Also, the exciton reservoir is emptied on a nano-second time-scale, typically. As a consequence, the newly created condensate, formed once the pump is switched on again,
completely loses the information about its previous state. If the
persistent circular current had arisen stochastically, as a result
of a chaotic uncertainty of the polariton density near the
condensation threshold~\cite{Ohadi2015,Ostr16,Nalitov2017},
then, the subsequent switching on of the excitation light would
have created interferential spirals of opposite helicities with
equal probabilities. However, the experimentally observed spiral
conserves its helicity in every switching off and on of the
pump. Since, in subsequent switching on of the excitation
light all the experimental parameters remain unchanged, we can
conclude that the appearance of circular currents is most likely due to the combined effect of 
structural inhomogeneities of the microcavity and the microscopic shift of the pump spot from the center of the pillar. The combined effect of a stationary disorder and shifted pump spot is capable of breaking the symmetry between clockwise and anticlockwise polariton flows. In is important to note that, taken alone, the shift of the pump spot away from the center of the pillar would not break this symmetry. The stationary potential created by the pillar must not be perfectly cylindrically symmertic, that makes us thinkng of hidden structures inhomogeneities of the pillar. The nature of these
inhomogeneities is unknown. Because of the sharp dependence of the shape of our interferograms on microscopic pumping spot displacements, we believe that most likely the stationary disorder comes from a local defect situated near the pillar center, which breaks the axial
symmetry of scattering of polaritons sliding down the potential
hill formed by hot excitons under the excitation spot.

\begin{figure}[t]
\includegraphics[width=\linewidth]{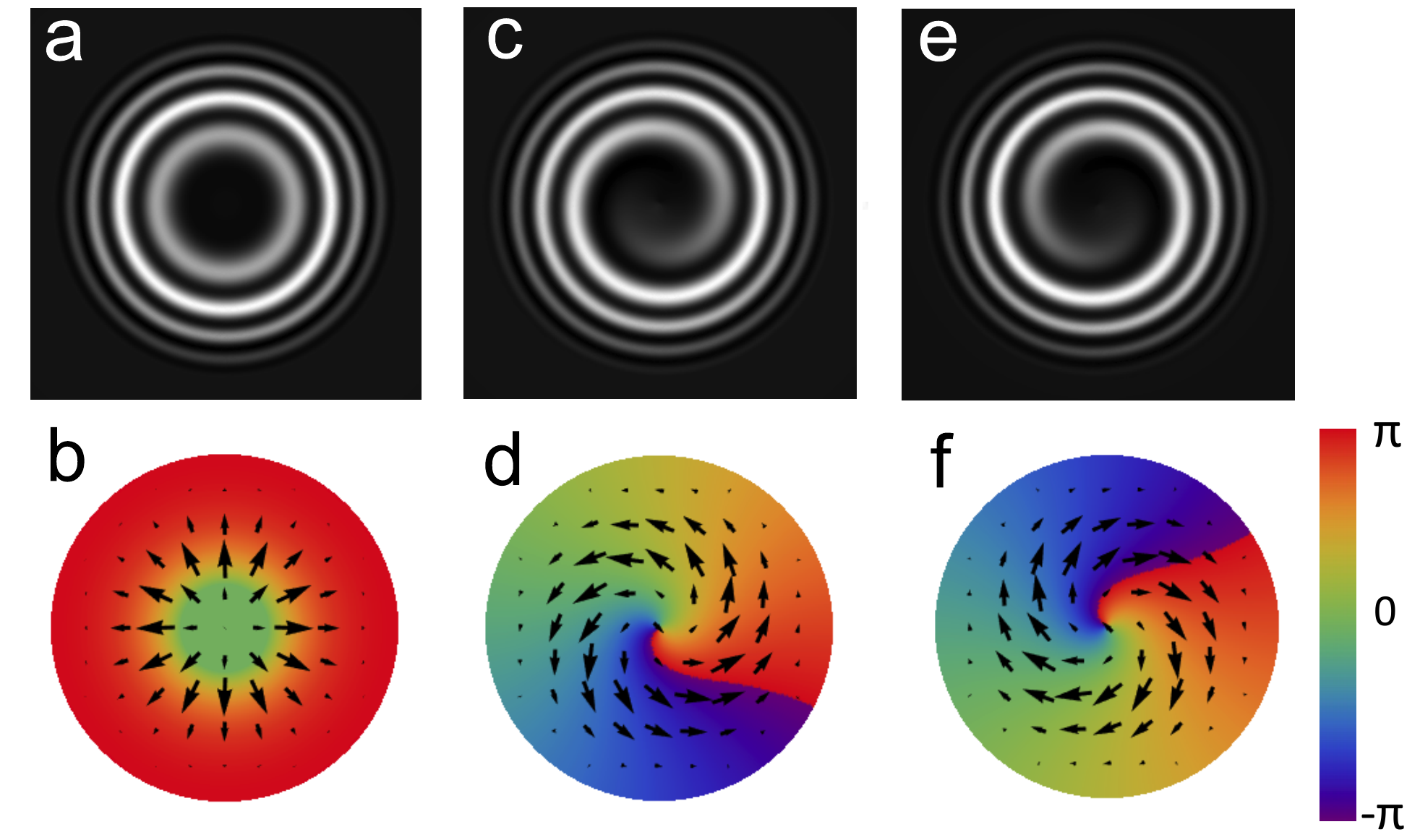}
\caption{ \label{Fig3} (Color online) The interference pattern (upper panels) and  the spatial distribution of the condensate phase (lower panels)
calculated for different values of the topological charge: $m=0$
for (a) and (b), $m=+1$ for (c) and (d), and $m=-1$ for (e) and
(f). Black arrows in the phase maps (b), (d) and (f) denote the
vector field $\mathbf{J} = \text{Im} (\psi ^{*} \nabla \psi)$. The
parameters used for the modelling are given in
Ref.~\cite{BibModellingParameters}. }
\end{figure}

Within the mean-field approach, the experimentally observed
exciton-polariton condensate states can be approximated by the
azimuthally symmetric stationary condensate wavefunction which we
search in the form $\Psi (\mathbf{r},t) = \psi(r) \exp (i m
\theta) \exp(-i \mu t)$ (see Refs.\,\cite{PRA860136362012,Ostr}),
where $\mu$ is the chemical potential,
%and $m$ is the topological charge of a vortex~\cite{DESYATNIKOV2005291,PRA860136362012,PRB911845182015};
$r$ and $\theta$ are the polar coordinates.
%The topological charge characterizes the normalised angular momentum, $m = i \left.\int ( \psi \partial _{\theta} \psi ^{*} - \psi ^{*}\partial _{\theta} \psi ) r dr  \right/ 2 \int |\psi|^2 r dr  $, which in the rotationally symmetric problem is the conserved quantity
The wavefunction $\psi$ obeys the stationary generalized Gross-Pitaevskii equation
\begin{equation}
\label{StGPE}
\mu \psi = \left [
\widehat{E}_{\text{kin}}
+ V_{\text{eff}} (r)
- \left. i \hbar \left( \gamma _C - R n_{R}(r)  \right)  \right /2
 \right]  \psi,
\end{equation}
where $\widehat{E}_{\text{kin}} = \left. -\hbar^2 \nabla ^2
\right/ 2 M $ is the kinetic energy operator, $M$ is the polariton effective
mass, the operator $\nabla ^2$ in the polar
coordinates is $\nabla ^2 = \partial _{r r} +r^{-1}
\partial _r - m ^2 r^{-2}$. The stationary effective potential
$V_{\text{eff}}(r)$ takes the form $V_{\text{eff}} (r) = V (r) +
\alpha _C |\psi|^2 + \alpha _R n_{R}(r)$. The term $V(r)$
describes the stationary trapping potential governed by the geometry of the structure. For the cylindrical pillar of a
diameter $d$ it is given by $V (r) = V_0 \Theta (r - d/2) $, where
$V_0$ is the height of the potential and $\Theta (r - d/2)$ is the
Heaviside step function. The second and the third terms in
$V_{\text{eff}}(r)$ describe corrections to the potential
landscape due to the intra-condensate polariton-polariton
interactions and the polariton interactions with the reservoir of
hot excitons, respectively. The parameters $\alpha _C$ and $\alpha
_R$ are the corresponding polariton-polariton and
polariton-exciton coupling constants. The stationary density of excitons in the reservoir  $n_{R}(r) = \left. P (r) \right/ (\gamma_R + R |\psi|^2 )$,
where $P(r)$~is the spatially inhomogeneous nonresonant optical
pump~\cite{BibPumpDetails} and $R$ is the stimulated scattering
rate describing particle exchange between the polariton condensate
and the exciton reservoir. The imaginary term in the right-hand
side of Eq.\,\eqref{StGPE} is responsible for the balance of the
gain from the pumped reservoir and the losses due to the finite polariton lifetime. The factors $\gamma_C$ and $\gamma_R$ are the loss rates of the condensate polaritons and the
reservoir excitons, respectively.

The upper panels in
Fig.~\ref{Fig3} demonstrate the results of the numerical simulation
of the stationary intensity patterns appearing due to the interference of light emitted by the ring-shaped polariton condensates with different topological charges, $m=0$, $+1$ and
$-1$ for (a), (b) and (c), respectively, with the spherical wave~\cite{SphericalWave}. In the modelling, the effect of the structural inhomogeneities of the microcavity resulting in the condensate circular
currents is taken into account by choosing the absolute value and the sign of the winding number $m$. The simulated interferograms for $m = ± 1$ have the shape of Fermat spirals, with the fringes obeying the relation $r^2 \propto \theta$, in agreement with experimental patterns shown in Fig.\,\ref{Fig2}c,d. In the  case of $m=0$, the calculated interference pattern takes the form of a set of concentric rings reproducing the experimental image in Fig.\,\ref{Fig2}b. The corresponding spatial distribution
of the condensate phase toghether with the vector field
$\mathbf{J} = \text{Im} (\psi ^{*} \nabla \psi)$ characterizing
the superfluid polariton flow in the stationary state are shown in the
lower panels in Fig.\,\ref{Fig3}. The topological
charge $m$ and the polariton flow are linked to each other directly by the expression $m = (2 \pi N)^{-1} \int _{S}
(\partial _x J_y -
\partial _y J_x) d \mathbf{r}$, where $S$ is the area enclosing
the condensate density distribution~\cite{Ostr16} and $N$ is the
wavefunction normalization constant describing the population of
the condensate. In Figs.\,3d,f the vector field $\mathbf{J}$ winds 
around the vortex core. It directly shows the direction and the magnitude of the superfluid polariton current. The spirals in interferograms correspond to the persistent current states, while the interferogram with concetric rings is characteristic for the azimutally symmetric phase distribution (no current). 

\begin{figure}[t]
\includegraphics[width=\linewidth]{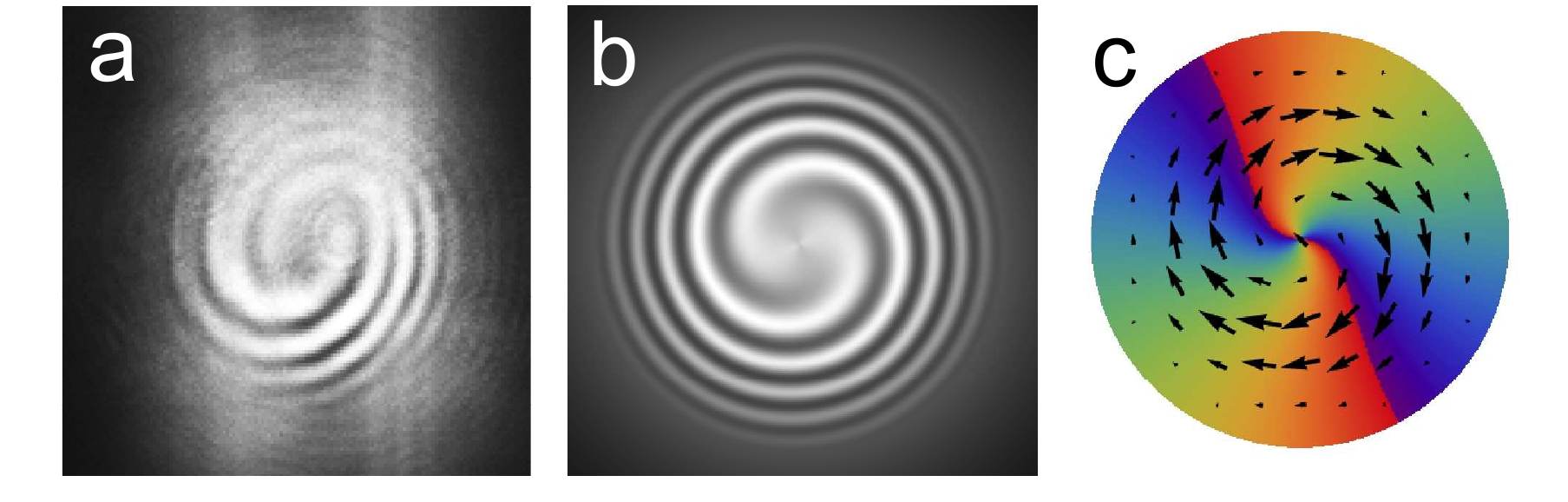}
\caption{
\label{Fig4} (Color online)
The experimental interferogram of the exciton-polariton condensate with the topological charge $ m = -2$ recorded in 25\,$\mu$m cylindrical pillar (a).
The modeled interferogram (b) and real space distrubution of the phase (c) of the $m=-2$ condensate. Arrows in map (c)  show the direction and the magnitude of the superfluid flow $\mathbf{J}$ of polaritons. The color scale in panel (c) is the same as in Figs.\,3b,d,f.
}
\end{figure}

While persistent currents with $m = ± 1$ are relatively easy to obtain, higher topological charges can scarcely be seen. So far, we managed to experimentally realize only the state with $m=-2$.
The corresponding interferogram given in Fig.\,\ref{Fig4}a
represents twin spirals emerging from the pillar center clockwise.
The twin-spiral state, once obtained, also demonstrates a high
stability, persisting for a few minutes. It was modelled in the
same way as the states with topological charges $+1$ and $-1$. The corresponding stationary interference
pattern and the spatial distribution of the phase are shown in
Figs.\,\ref{Fig4}b and \ref{Fig4}c, respectively.

In conclusion, in this Letter we have demonstrated experimentally
the existence of persistent circular currents in the ring-shaped
exciton-polariton condensates appearing in cylindrical micropillars
under the nonresonant optical pumping. We have observed the quantum states of ring-shaped condensates characterised by topological charges $m = ± 1, -2$. The superfluid polariton currents are sustained by the balance of the gain due to the optical pump and the loss
due to the finite polariton lifetime. Once established, the
current is preserved by the condensate within the time duration of
the experiment. Its topological charge would not change at
switching the optical pumping off and on. This observation indicates that the symmetry breaking between clockwise and anticlockwise polariton flows is a result of the interplay of an asymmetric stationary disorder potential and the microscopic shift of the pump spot from the center of the pillar. In spite of the yet unknown nature of the disorder imperfections leading to the vorticity, cylindrical micropillars proved to be convenient
objects for realization of circular polariton currents. The observation of persistent circular currents of exciton-polaritons paves the way to multiple applications of polariton condensates in quantum interference devices.

\acknowledgements The work of VAL, VKK, MMA, KVK and AVK 
was supported in part by the RFBR (Grant No. 15-52-12018) within the joint Russian–German project ICRC TRR160. 
PGS acknowledges Funding from the POLAFLOW ERC Starting Grant.
ESS thanks the RFBR Grant No. 16-32-60104.
AVK acknowledges the support from the EPSRC Programme grant on Hybrid Polaritonics No. EP/M025330/1 and the partial support from the HORIZON 2020 RISE project CoExAn (Grant No. 644076).


\begin{thebibliography}{99}
\bibitem{BibPumpDetails} We consider the azimuthally symmetric Gaussian pump, $P(r) = P_0 \exp (- \left. r^2 \right/ w_p^2 )$, centered
on the pillar.  The intensity $P_0$ and the width of the pump beam
$w_p$ are the fitting parameters in the modelling.
\bibitem{BibModellingParameters} In the modelling, we take the following values of the parameters: the effective mass of polaritons is ${M = 5 \cdot 10^{-5} m_e}$ with $m_e$ being a free electron mass,
the decay rates are $\gamma _C = 0.05 \, \text{ps}^{-1}$ and $\gamma _R = 10 \gamma _C$, the interaction constants are $\alpha _C = 2\, \text{meV}\,\mu\text{m} ^2 $ and $\alpha _R = 10 \alpha _C$, the condensate-reservoir scattering rate is $\hbar R = 0.05\, \text{meV} \,\mu\text{m} ^2 $.
\bibitem{SphericalWave}
The well-known expression for the spherical wave is ${E_{S}
\propto \left. \exp(i k \rho) \right/ \rho}$, where $k$ is the
light wavenumber, ${\rho = \sqrt{ r^2 + l^2}}$ is the radius of
the wavefront, $l$ is the shortest distance from the center of the
spherical wave to the plane of the monochromator entrance slit,
which is the fitting parameter in the modelling.
\end{thebibliography}
\end{document}